\documentclass[aps,prc,twocolumn,superscriptaddress,showpacs]{revtex4}
\usepackage{graphicx}
\usepackage{bm}

\begin{document}

\title[Neutral Koan Interferometry at STAR]{Neutral Kaon Interferometry
in Au+Au collisions at $\sqrt{s_{NN}} = 200$ GeV}

\date{\today}

\affiliation{Argonne National Laboratory, Argonne, Illinois 60439}
\affiliation{University of Birmingham, Birmingham, United Kingdom}
\affiliation{Brookhaven National Laboratory, Upton, New York 11973}
\affiliation{California Institute of Technology, Pasadena, California 91125}
\affiliation{University of California, Berkeley, California 94720}
\affiliation{University of California, Davis, California 95616}
\affiliation{University of California, Los Angeles, California 90095}
\affiliation{Carnegie Mellon University, Pittsburgh, Pennsylvania 15213}
\affiliation{University of Illinois, Chicago}
\affiliation{Creighton University, Omaha, Nebraska 68178}
\affiliation{Nuclear Physics Institute AS CR, 250 68 \v{R}e\v{z}/Prague, Czech Republic}
\affiliation{Laboratory for High Energy (JINR), Dubna, Russia}
\affiliation{Particle Physics Laboratory (JINR), Dubna, Russia}
\affiliation{University of Frankfurt, Frankfurt, Germany}
\affiliation{Institute of Physics, Bhubaneswar 751005, India}
\affiliation{Indian Institute of Technology, Mumbai, India}
\affiliation{Indiana University, Bloomington, Indiana 47408}
\affiliation{Institut de Recherches Subatomiques, Strasbourg, France}
\affiliation{University of Jammu, Jammu 180001, India}
\affiliation{Kent State University, Kent, Ohio 44242}
\affiliation{Institute of Modern Physics, Lanzhou, China}
\affiliation{Lawrence Berkeley National Laboratory, Berkeley, California 94720}
\affiliation{Massachusetts Institute of Technology, Cambridge, MA 02139-4307}
\affiliation{Max-Planck-Institut f\"ur Physik, Munich, Germany}
\affiliation{Michigan State University, East Lansing, Michigan 48824}
\affiliation{Moscow Engineering Physics Institute, Moscow Russia}
\affiliation{City College of New York, New York City, New York 10031}
\affiliation{NIKHEF and Utrecht University, Amsterdam, The Netherlands}
\affiliation{Ohio State University, Columbus, Ohio 43210}
\affiliation{Panjab University, Chandigarh 160014, India}
\affiliation{Pennsylvania State University, University Park, Pennsylvania 16802}
\affiliation{Institute of High Energy Physics, Protvino, Russia}
\affiliation{Purdue University, West Lafayette, Indiana 47907}
\affiliation{Pusan National University, Pusan, Republic of Korea}
\affiliation{University of Rajasthan, Jaipur 302004, India}
\affiliation{Rice University, Houston, Texas 77251}
\affiliation{Universidade de Sao Paulo, Sao Paulo, Brazil}
\affiliation{University of Science \& Technology of China, Hefei 230026, China}
\affiliation{Shanghai Institute of Applied Physics, Shanghai 201800, China}
\affiliation{SUBATECH, Nantes, France}
\affiliation{Texas A\&M University, College Station, Texas 77843}
\affiliation{University of Texas, Austin, Texas 78712}
\affiliation{Tsinghua University, Beijing 100084, China}
\affiliation{Valparaiso University, Valparaiso, Indiana 46383}
\affiliation{Variable Energy Cyclotron Centre, Kolkata 700064, India}
\affiliation{Warsaw University of Technology, Warsaw, Poland}
\affiliation{University of Washington, Seattle, Washington 98195}
\affiliation{Wayne State University, Detroit, Michigan 48201}
\affiliation{Institute of Particle Physics, CCNU (HZNU), Wuhan 430079, China}
\affiliation{Yale University, New Haven, Connecticut 06520}
\affiliation{University of Zagreb, Zagreb, HR-10002, Croatia}

\author{B.I.~Abelev}\affiliation{Yale University, New Haven, Connecticut 06520}
\author{M.M.~Aggarwal}\affiliation{Panjab University, Chandigarh 160014, India}
\author{Z.~Ahammed}\affiliation{Variable Energy Cyclotron Centre, Kolkata 700064, India}
\author{J.~Amonett}\affiliation{Kent State University, Kent, Ohio 44242}
\author{B.D.~Anderson}\affiliation{Kent State University, Kent, Ohio 44242}
\author{M.~Anderson}\affiliation{University of California, Davis, California 95616}
\author{D.~Arkhipkin}\affiliation{Particle Physics Laboratory (JINR), Dubna, Russia}
\author{G.S.~Averichev}\affiliation{Laboratory for High Energy (JINR), Dubna, Russia}
\author{Y.~Bai}\affiliation{NIKHEF and Utrecht University, Amsterdam, The Netherlands}
\author{J.~Balewski}\affiliation{Indiana University, Bloomington, Indiana 47408}
% ...

\collaboration{STAR Collaboration}\noaffiliation

\begin{abstract}

We present the first statistically meaningful results from
two-$K^0_s$ interferometry in heavy-ion collisions. A model that
takes the effect of the strong interaction into account has been used to
 fit the measured correlation function. The effects of single and coupled
channel were explored. At the mean transverse mass $\langle m_T \rangle$ = 1.07 GeV, 
we obtain the values $R = 4.09 \pm 0.46 (stat.) \pm 0.31 (sys)$ fm
and $\lambda = 0.92 \pm 0.23 (stat) \pm 0.13 (sys)$, where $R$ and $\lambda$ 
are the invariant radius and chaoticity parameters respectively. The results are 
qualitatively consistent with $m_T$ systematics established with pions 
in a scenario characterized by a strong collective flow.
\end{abstract}


\maketitle

\section{Introduction}
\newcounter{mycounter1}[section]

Lattice QCD calculations predict that a phase transition from hadronic matter to
a new state of matter called a Quark Gluon Plasma (QGP) occurs at sufficiently
 large energy densities \cite{LATQCD}. Creation and study of such a de-confined
 state of matter is the primary goal of the heavy-ion collisions program at the
 Relativistic Heavy-Ion Collider (RHIC). A first order phase transition from the QGP back
 to normal hadronic matter is believed to delay the expansion of the hot reaction
 zone created in the collision \cite{PRATT1}. A delayed expansion means a
 long duration of particle emission, leading to a large source size.

The measurement of the space-time extent of the particle emitting
region has been one of the important goals in high energy
experiments for several decades \cite{MIPOD,UAWIED,RLED}. These
measurements are based on the sensitivity of particle  momentum
correlations to the space-time separation of the particle
 emitters due to the effects of quantum statistics (QS) and final state
interaction(FSI). For identical particles, the QS symmetrization (antisymmetrization)
is usually the dominant source of the correlation and, due to the interference of the
amplitudes corresponding to various permutations of identical particles, this
measurement is often called particle interferometry (See reference \cite{MLISA} for a review).

Most of the particles produced in relativistic heavy-ion collisions
are pions and, as a result, pion interferometry has been a
particularly useful tool in correlation studies. High statistics
data from colliders like RHIC have also made it possible to study
kaon correlations.
 In this Letter, we present the first results on two-$K^0_s$ correlations in 
central Au-Au collisions at $\sqrt{s_{NN}} = 200$ GeV measured by the
STAR(Solenoidal Tracker at RHIC) experiment at RHIC .

It is known that a significant fraction of pions come from resonance decays 
after freeze-out thus complicating the pion interferometry measurements.
While the direct pion source may be inherently non-Gaussian,
 the resonances extend the source size due to their finite lifetime,
introduce an additional essentially non-Gaussian distortion in
the two-pion correlator and reduce the fitted correlation strength.
Due to the limited decay momenta, the decay pions populate mainly
the low momentum region, thus introducing an additional
 pair momentum dependence in the correlator.

 Kaon interferometry, on the other hand, suffers less from resonance
contributions and could provide a cleaner signal for correlation
studies than pions \cite{GYULASSY1, SULLIVAN1}. Higher multi-particle
 correlation effects, that might play a role for pions, should be of
 minor importance for kaons since the kaon density is considerably
 smaller than the pion density at RHIC ($\sqrt{s}_{NN} = 200$ GeV).
 The pion multiplicity has increased by approximately $70\%$ from
the SPS($\sqrt{s}_{NN} = 17.3$ GeV) to RHIC \cite{PHOBOS1}. The
interferometry radii however remain almost the same \cite{STAR1,PHENIX1}.
 The strangeness distillation mechanism \cite{GREINER1} might further
 increase any time delay QGP signature. This mechanism could lead to
strong temporal emission asymmetries between kaons and anti-kaons
\cite{SOFF1},
 thus probing the latent heat of the phase transition.

Particle identification for pions, via their specific ionization
 (energy loss per unit length or $dE/dx$), works only up to about 700 MeV/$c$.
 Neutral kaons, on the other hand, can be identified up to much higher
momentum using their decay topology. This allows for the extension
of the interferometry systematics to a higher momentum than is
presently achievable with pions, and thus provide a means to probe
the earlier times of the collision. The effect of two-track resolution, which is a limiting factor
 in charged particle correlations, is also small. The absence of Coulomb FSI suppression together
with small contributions from resonance decays make neutral kaon correlations
 a powerful tool to investigate the space time structure of the particle
emitting source.

The OPAL \cite{OPAL00} and ALEPH \cite{ALEPH94} collaborations have measured
correlations of neutral kaons from hadronic decays of $Z^0$ in $e^+e^-$
collisions at LEP. The WA97 experiment at CERN \cite{WA97_CERN} attempted to measure
$K^0_sK^0_s$ correlations but did not see a significant enhancement of neutral
 kaon pairs in the region of small momentum difference due to a lack of sufficient
 statistics.

\section{The STAR Experiment}

The STAR detector \cite{STARDET} consists of several detector
subsystems in a large solenoidal magnet that provides a uniform 0.5
Tesla field. For the data used
 in this analysis, the main setup consisted of the time projection chamber
(TPC) \cite{STARTPC} for charged particle tracking, a scintillator trigger
barrel (CTB) surrounding the TPC for measuring charged particle  multiplicity,
 and two zero degree calorimeters (ZDC) \cite{STARZDC} located upstream and
downstream along the axis of the TPC and beams to detect spectator
neutrons.
 With full azimuthal coverage over $|\eta| < 1$ and an almost $100\%$ efficiency
 for minimum ionizing particles, the CTB provides a good estimate of the number
 of charged particles produced in the mid-rapidity region. The number of neutrons
detected in the ZDC's is identified with the amount of energy deposited in them.
 The collision centrality is determined by correlating the energy deposition in
 the ZDC  with the number of minimum ionizing particles detected by the CTB.

\begin{figure}[h]
\centering
\begin{minipage}[h!]{0.48\textwidth}
\centering
\includegraphics[width=\textwidth,height=0.8\textwidth]{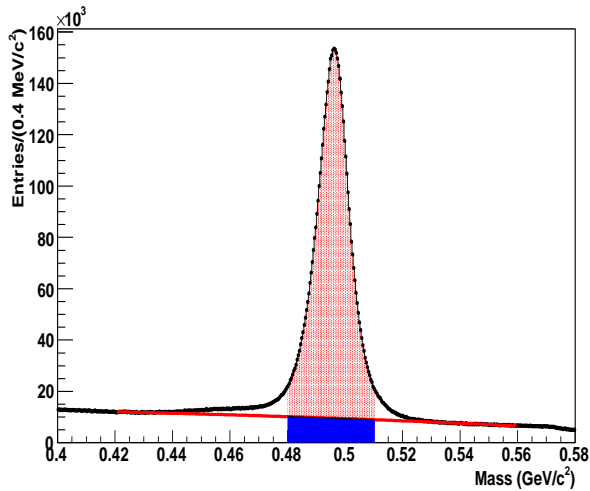}
\end{minipage}\hfill
\caption{The $K^0_s$ invariant mass distribution. The range in
transverse momentum is from $0.5$ GeV/$c$ to $3.5$ GeV/$c$ and
rapidity is between -1.5 and 1.5. Kaon candidates falling in the
mass range from $0.48$ GeV/$c^2$ to $0.51$ GeV/$c^2$, indicated 
by the shaded region, were selected for this correlation study.
The corresponding mass is $495.6 \pm 6.8$ MeV/$c^2$.\label{K0sMassPeak}}
\end{figure}

\subsection{Data Selection}

For this analysis, events from the ZDC central trigger ($0-10\%$ of the total hadronic
cross section) were used with an event vertex within $\pm 25$ cm of the center of the TPC,
 along the beam axis. Approximately $2.5 \times 10^6$ events with about 3 $K^0_s$
per event on the average were analyzed. Here we discuss $K^0_s$-specific issues only, as
 details of pion interferometry at the STAR experiment have been discussed in
\cite{MERCEDES}. The $K^0_s$ has a mean decay length ($c\tau$) of $2.7$~cm and decays via 
the weak interaction into $\pi^+$ and $\pi^-$ with a branching ratio of about $68\%$. 
The mass and kinematic properties of the $K^0_s$ are determined from the decay vertex 
geometry and daughter particle kinematics \cite{SBPHDTHESIS}. Neutral kaon candidates are
 formed out of a pair of  positive and negative tracks whose trajectories  point to a common
 secondary decay vertex which is well separated from the primary event vertex.
 All neutral kaon candidates, with invariant masses from $0.48$ GeV/$c^2$ 
to $0.51$ GeV/$c^2$, transverse momentum from $0.5$ GeV/$c$ to $3.5$ GeV/$c$ and rapidity
 between -1.5 and 1.5 have been considered. The daughter particle tracks are required to
 have a minimum of 15 TPC hits and a distance of closest approach to the primary vertex 
greater than 1.3 cm.

\subsection{The Correlation function}

Experimentally, the two-particle correlation function is defined as

\begin{equation}
C_2(Q) = \frac{A(Q)}{B(Q)} ,
\end{equation}
where $A(Q)$ represents the distribution of the
invariant relative momentum $Q = \sqrt{-q^{\mu}q_{\mu}}, q^{\mu}=p^{\mu}_{1}-p^{\mu}_{2}$,
for a pair of particles from
the same event. The possibility of a single neutral kaon being correlated with itself,
i.e., correlation between a real $K^0_s$ and a fake $K^0_s$ reconstructed from a pair which shares
 a daughter of the real $K^0_s$, was eliminated by requiring that kaons in a pair have unique daughters.
 We have also explored effects from splitting of daughter tracks by looking at the angular correlation
 between the normal vectors to the decay planes of the $K^0_s$ in a given pair. No
enhancement at very small angles was observed indicating no significant problem from
 track splitting. $B(Q)$ is the reference distribution constructed by mixing particles
from different events with similar Z-vertex positions(relative z position within 5 cm).
  The individual $K^0_s$ for a given mixed pair are required to pass the same single 
particle cuts applied to those that go into the real pairs. The mixed pairs are also 
required to satisfy the same pairwise cuts applied to the real pairs from one event. 
The efficiency and acceptance effects cancel out in the ratio $\frac{A(Q)}{B(Q)}$.

\subsection{Data Analysis}

 Figure \ref{K0sMassPeak} shows the invariant mass distribution of the neutral kaons
 based on the set of cuts described above. The background is characterized by a polynomial
 fit to the distribution outside the mass peak. The observed mass $495.6 \pm 6.8$ MeV/$c^2$
is consistent with the accepted value \cite{PDG}. The signal and background for the mass range
from $0.48$ ~GeV/$c^2$ to $0.51$ ~GeV/$c^2$ considered in this analysis are shown by the
 shaded regions. 

\begin{figure}[h]
\centering
\begin{minipage}[h!]{0.48\textwidth}
\centering
\includegraphics[width=\textwidth,height=0.8\textwidth]{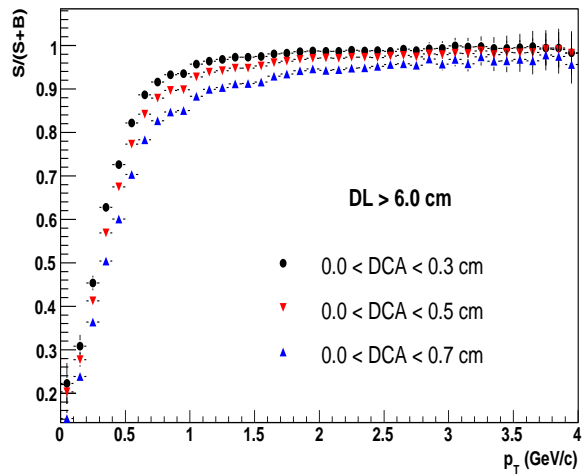}
\end{minipage}\hfill
\caption{The $K^0_s$ signal to (signal+background) ratio as a function of the
transverse momentum $p_T$. The data points correspond to a decay
length (DL) greater than 6 cm. The kaons selected fall in the mass
range from $0.48$ ~GeV/$c^2$ to $0.51$ ~GeV/$c^2$ which is also the
 mass range for the correlation analysis. The errors are only statistcal.
 \label{SIGTONOISE}}
\end{figure}
\begin{figure}[h]
\centering
\begin{minipage}[h!]{0.52\textwidth}
\centering
\includegraphics[width=\textwidth,height=1.2\textwidth]{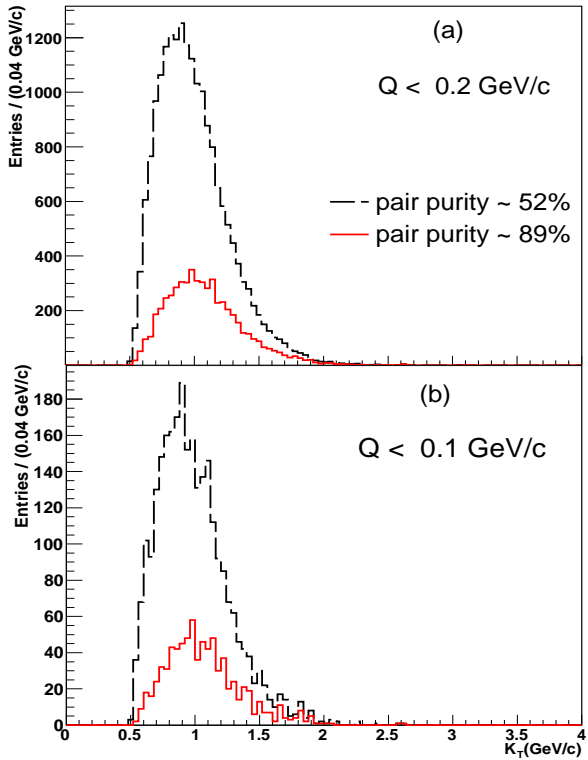}
\end{minipage}\hfill
\caption{The $K_T$ distribution of the $K^0_s$ pairs. The
range in transverse momentum of the single particles is from $0.5$
~GeV/$c$ to $3.5$ ~GeV/$c$. The distribution in (a) corresponds to
$Q < 0.2$ GeV/c and that in (b) is for $Q < 0.1$ GeV/c, i.e., (b) is
a subset of (a). The two histograms in each panel are for low (dashed)
 and high (full) pair purity.'  \label{K0sKtDist}}
\end{figure}

After tuning several kinematical and detector related cuts to remove most of the
background, some residual noise still remains. This calls for a knowledge of the
signal to background ratio within the selected invariant mass range to make corrections
 to the measured correlation function. For neutral kaons, the decay length (DL) and
distance of closest approach (DCA) to the interaction vertex were
two of the parameters for which it was difficult to determine where
to apply the cuts. Various DCA and DL cut combinations  were
investigated by varying the DCA from 0.3 cm to 0.8 cm in steps of 0.1
cm and the DL from 2.0 cm to 6.0 cm in steps of 1.0 cm. Figure
\ref{SIGTONOISE} displays an example of the signal to background
ratio as a function of $p_T$ for DL $> 6 $ cm and various DCA
values. The single particle purity gets worse as the DCA gets larger
for the given DL cut. If one instead looks at a fixed DCA and varies
 the DL cut instead, the purity gets better with increasing decay
length.

The effect of momentum resolution on the correlation functions has also been investigated using
 simulated tracks from $K^0_s$ decays with known momenta, $\vec{p}_{in}$, embedded into real events.
 The reconstructed momenta of the embedded tracks, $\vec{p}_{rec}$, are then compared with
 $\vec{p}_{in}$. The distributions of $\frac{|\vec{p}_{rec} -
 \vec{p}_{in}|}{|\vec{p}_{in}|}$ with respect to $\vec{p}_{in}$
 are
 then fit to Gaussians to obtain the RMS widths. These are used to characterize the momentum
resolution of the detector. The resolution in $p$ lies between $1\%$
and $2\%$ for the $p_T$ range used in this analysis.

 The top panel in Figure \ref{K0sKtDist} shows the $K_T$ distribution for $Q < 0.2$
GeV/c where $K_T =(|\vec{p}_{1T} + \vec{p}_{2T}|)/2$. The correponding number of pairs for
 the distribution with low pair purity is approximately $1.92 \times 10^4$ and that for the one with
the high pair purity is about $5.5 \times 10^3$. The distribution in the bottom panel corresponds to
 pairs with $Q < 0.1$ GeV/c, with  $2.7 \times 10^3$ pairs for the low pair purity distribution
and  $7.8 \times 10^2$ for the high pair purity distribution.
It is clear that the shape of the $K_T$ distribution changes with the pair purity and, as a result, so does
$\langle K_T \rangle$, the mean of the distribution. The mean $K_T$ varies almost linearly with pair purity.
For the lowest pair purity value
 of $\approx 52\%$, $\langle K_T \rangle \approx  0.805 $ GeV/c. At the highest pair
 purity value of $\approx 89\%$, $\langle K_T \rangle \approx 1.07 $ GeV/c. The dependence of $\langle K_T \rangle$
on the pair purity together with the fact that the radii may vary
with $K_T$ implies that varying the pair
 purity may change the measured radii. In this analysis, the correlation function is integrated
 over all $K_T$ since the statistics are not sufficient to make a $K_T$ dependent study.

Corrections to the raw correlation functions were applied  according to the expression
\begin{equation}
        C_{corrected}(Q) = \frac {C_{measured}(Q) - 1}{PairPurity(Q)} + 1
\end{equation}
where the pair purity was calculated as the product of the signal($S$) to signal plus background
 ($S + B$) ratios of the two $K^0_s$ of the pair (i,j)
\begin{equation}
        PairPurity(Q) = \frac{S}{S + B}(p_{ti}) \times \frac{S}{S + B}(p_{tj})
\end{equation}

 The pair purity, $PairPurity(Q)$, has been found to be independent of $Q$ over the range of
invariant four-momentum difference considered. As a result, an average value over Q of the 
pair purity has been used to correct the correlation function for each set of cuts considered.

\begin{figure}[ht]
\centering
\begin{minipage}[h!]{0.48\textwidth}
\centering
\includegraphics[width=\textwidth,height=0.8\textwidth]{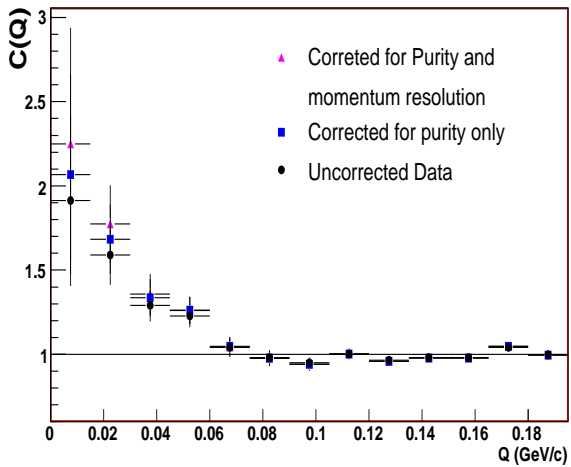}
\end{minipage}\hfill
\caption{The $K^0_sK^0_s$ correlation function. The solid circles
are for uncorrected data. The squares correspond to the case where
the data have been corrected for pair purity. The triangles
represent the data after correcting for pair purity and momentum
resolution. The errors are ststistical only. \label{CorrectedCF}}
\end{figure}

Figure \ref{CorrectedCF} shows the experimental $K^0_sK^0_s$ correlation function before and after
corrections for purity and momentum resolution are applied. It can be seen that the effect of momentum
 resolution is comparable to that of purity correction.
The one-dimensional correlation function is usually fitted to a
Gaussian
\begin{equation}
C(Q)= N\cdot({1 + \lambda\cdot{e^{-R^2Q^2}}})
\label{K0CorrEq}
\end{equation}
where $N$ and $R$ are respectively the normalization and size parameter, the latter
characterizing the width of the Gaussian distribution of the vector
$\vec{r}^*$
of the relative distance between particle emission points in the pair c.m.s.:
\begin{equation}
\frac{d^3N}{d^3\vec{r}^*} \propto e^{-\vec{r}^{*2}/(4R^2)}.
\label{rDIST}
\end{equation}
The parameter $\lambda$ measures the correlation strength. In the
absence of FSI, $\lambda$
 equals unity for a fully chaotic Gaussian source, up to a suppression due to
the kaon impurity and finite momentum resolution. Theoretically, it can be less than
 unity due to partial coherence of the kaon field, resonance decays and the non-Gaussian
 form of the correlation function. Also neglecting FSI can affect (suppress or enhance)
 the value of this parameter.

\section{Final State Interaction In The Neutral Kaon System}

The production of the neutral kaon system, $K^0$ and $\bar{K^0}$,
 is attributed to the strong interaction which conserves the strangeness
quantum number. An interesting property of neutral kaons is that the
$K^0$
 can change into a $\bar{K^0}$ through a second order weak interaction. However,
 the particles that we normally observe through weak decay channels in the
laboratory are not $K^0$ and $\bar{K^0}$ \cite{GELLMANN1}. Neglecting the effects
 of CP violation, the observed weak interaction eigenstates are given by  

\begin{eqnarray}
|K^0_s\big \rangle &=& \frac{1}{\sqrt{2}}(|K^0\big \rangle + |\bar{K^0}\big \rangle) ,\nonumber\\
|K^0_l\big \rangle &=& \frac{1}{\sqrt{2}}(|K^0\big \rangle - |\bar{K^0}\big \rangle) ,
\end{eqnarray}
where $|K^0_s\big \rangle$ and $|K^0_l\big \rangle$ are the state
vectors of the short and long lived neutral kaons, to which
experiments have access via measurements of their decay products, 
which are mainly pions.
The state vector of the $K^0_sK^0_s$ system is then given by the expression

\begin{eqnarray}
|K^0_sK^0_s \big \rangle &=& \frac{1}{2}(|K^0K^0\big \rangle + |K^0\bar{K^0}\big \rangle \nonumber\\
& & + |\bar{K^0}K^0\big \rangle + |\bar{K^0}\bar{K^0}\big \rangle).
\end{eqnarray}
Now, if a $K^0_sK^0_s$ pair comes from $K^0$$K^0$ ($\bar{K^0}$
$\bar{K^0}$), it is subject to Bose-Einstein (BE) enhancement as it
originates from an identical boson pair. On the other hand, the
$K^0$ and $\bar{K^0}$ are two
 different particles and one may not expect correlations if one $K^0_s$
comes from $K^0$ and the other one from $\bar{K^0}$. Nevertheless, it can
 be shown \cite{VLLYMIPOD} (see also \cite{ALEXANDER99,ALEXANDER1,GYULASSY2}) that
only the symmetric part of the $K^0\bar{K^0}$ amplitude contributes to
the $K^0_sK^0_s$ system and thus also leads to a Bose-Einstein
enhancement at small relative momentum (on the contrary, only the
anti-symmetric part of the $K^0\bar{K^0}$ amplitude contributes to the $K^0_sK^0_l$
 system and leads to the ``Fermi-Dirac like'' suppression). The $K^0_sK^0_s$ correlation
thus includes a unique interference term that may  provide additional
space-time information. Here only the $K^0_sK^0_s$ correlation is considered 
since most of the $K^0_l$ decay outside the STAR TPC and are not accessible.

The strong FSI has an important effect on neutral kaon correlations due to the near
threshold resonances, $f_{0}(980)$ and $a_{0}(980)$
\cite{LEDNICKY}. These resonances contribute to the $K^0\bar{K^0}$ channel and lead
to the s-wave scattering length dominated by the imaginary part of $\sim$1 fm. Based on the predictions of chiral
perturbation theory for pions \cite{BGJH} the non-resonant s-wave scattering lengths are expected to be $\sim$0.1 fm
 for both $K^0\bar{K^0}$ and $K^0K^0$ channels and can be neglected to a first approximation.

To calculate the $K^0_sK^0_s$ correlation function, we assume
$K^0$'s and $\bar{K^0}$'s emitted by independent single-kaon sources
so that the fraction of $K^0_sK^0_s$ pairs originating from
$K^0\bar{K}^0$ system is $\alpha=(1-\epsilon^2)/2$, where $\epsilon$
is the $K^0$-$\bar{K}^0$ abundance asymmetry. We have put
$\alpha=1/2$ based on the negligible $K^+$-$K^-$ abundance asymmetry
of $0.018\pm 0.106$ as measured under the same conditions by the
 STAR experiment \cite{STAR2}.
 The correlation function is calculated as a mixture
 of the average squares of the properly symmetrized $K^0K^0$,
 $\bar{K^0}\bar{K^0}$ and non-symmetrized ${K^0}\bar{K^0}$ wave
 functions, weighted by the respective $K^0_sK^0_s$ fractions.
To average over the relative distance vector $\vec{r}^*$,
we use the Lednick\'y \& Lyuboshitz  analytical model \cite{LEDNICKY},
 assuming $\vec{r}^*$ is distributed according to Eq.~(\ref{rDIST}) with a Gaussian radius
$R$. The model assumes that the non-symmetrized wave functions
$\Psi_{-\vec{k}^*}(\vec{r}^*)$ describing the elastic transitions can be written as a superposition of the plane and
spherical waves, the latter being dominated by the s-wave,
\begin{equation}
\label{wf}
\Psi_{-\vec{k}^*}(\vec{r}^*)=e^{-i \vec{k}^*\vec{r}^*}+ f(k^*) \frac{e^{ik^*r^*}}{r^*},
\end{equation}
where
$\vec{k^*}\equiv\vec{Q}/2$ is the three-momentum of one of the kaons
in the pair rest frame
and $f(k^*)$
is the s-wave  scattering amplitude for a given system.
Neglecting the scattered waves for the $K^0K^0$ and $\bar{K^0}\bar{K^0}$ systems (the corresponding $f(k^*)=0$)
one obtains the following expression for the $K^0_sK^0_s$ correlation function \cite{LEDNICKY}:
\begin{eqnarray}
\label{cft}
C(Q) &=& 1 + e^{-Q^2R^2} + \alpha \left[\left|\frac{f(k^*)}{R}\right|^2\right. + \nonumber\\
& &\left.\frac{4\Re f(k^*)}{\sqrt\pi R}F_1(QR)- \frac{2\Im f(k^*)}{R}F_2(QR)\right],
\end{eqnarray}
where $F_1(z) = \int_0^z dx e^{x^2 - z^2}/z$ and $F_2(z) = (1-e^{-z^2})/z$.
The s-wave $K^0\bar{K^0}$ scattering amplitude $f(k^*)$ is dominated by the near threshold s-wave isoscalar and
isovector resonances $f_{0}(980)$ and $a_{0}(980)$ characterized by their masses $m_r$ and respective couplings
$\gamma_r$ and $\gamma'_r$ to the $K\bar{K}, \pi\pi$ and $K\bar{K}, \pi\eta$ channels. Associating the amplitudes
$f_I$ at isospin $I=0$ and $I=1$ with the resonances $r=f_0$ and $a_0$ respectively, one can write \cite{LEDNICKY,MARTIN}
\begin{equation}
\label{fk}
f(k^*) = [f_0(k^*) + f_1(k^*)]/2,
\end{equation}
\begin{equation}
\label{fik}
f_I(k^*) = \gamma_r/[m_r^2-s-i\gamma_rk^*-i\gamma'_rk'_r].
\end{equation}
Here $s = 4(m_K^2 + k^{*2})$ and $k'_r$ denotes the momentum in the second ($\pi\pi$ or $\pi\eta$)
channel with the corresponding partial width $\Gamma'_r = \gamma'_rk'_r/m_r$.

There is a great deal of uncertainty in the properties of these resonances due to insufficiently
accurate experimental data and the different approaches used in their analysis. Fortunately, the
dominant imaginary part of the scattering amplitude is basically determined by the ratios of the
$f_{0}K\bar{K}$ to $f_{0}\pi\pi$ and $a_{0}K\bar{K}$ to $a_{0}\pi\eta$
 couplings whose variation is rather small \cite{VBARU}.
In this paper we use the resonance masses and couplings from (a) Martin {\em et al.} \cite{MARTIN},
(b) Antonelli \cite{Antonelli}, (c) Achasov {\em et al.} \cite{Achasov2}, (d) Achasov {\em et al.} \cite{Achasov2}
 (see Table~\ref{table:TBHBT2}) to demonstrate the impact of their characteristic uncertainties
on the calculated correlation function.

\begin{table}[h]
\begin{center}
\begin{tabular}{|c|c|c|c|c|c|c|}
\hline
$Ref.$ &$m_{f_0}$& $\gamma_{f_0K\bar{K}}$  & $\gamma_{f_{0}\pi\pi}$ &$m_{a_0}$ & $\gamma_{a_0K\bar{K}}$  & $\gamma_{a_{0}\pi\eta}$\\ \hline
a &$ 0.978 $ & $ 0.792 $ & $ 0.199  $ & $ 0.974 $ & $ 0.333  $ & $ 0.222 $\\ \hline
b &$ 0.973 $ & $ 2.763 $ & $ 0.5283 $ & $ 0.985 $ & $ 0.4038 $ & $ 0.3711$ \\ \hline
c &$ 0.996 $ & $ 1.305 $ & $ 0.2684 $ & $ 0.992 $ & $ 0.5555 $ & $ 0.4401$ \\ \hline
d &$ 0.996 $ & $ 1.305 $ & $ 0.2684 $ & $ 1.003 $ & $ 0.8365 $ & $ 0.4580$ \\ \hline

\end{tabular}\\
\end{center}
\caption{The $f_0$ and $a_0$ masses and coupling parameters, all in GeV, from
(a) Martin {\em et al.} \cite{MARTIN},
(b) Antonelli {\em et al.}\cite{Antonelli},
(c) Achasov {\em et al.}\cite{Achasov2} and (d) Achasov {\em et al.} \cite{Achasov2}.
\label{table:TBHBT2}}
\end{table}
We have taken into account the normalization and correlation
strength parameters $N$ and $\lambda$ by the substitution $C(Q) \to
N\cdot [\lambda \cdot C(Q) + (1-\lambda)]$ . Following Ref.
\cite{LLL}, we have also included a small contribution of
the inelastic transition between the coupled channels $K^+K^- (\equiv 2)$
and $K^0\bar{K^0} (\equiv 1)$ (see Appendix for more details). Besides a
direct contribution of the average square of the corresponding wave
function $\Psi^{21}_{-\vec{k}^*}(\vec{r}^*)$ given in
Eq.~(\ref{wf21}), this transition also leads to a modification of
the amplitude $f(k^*)$ in the wave function of the elastic
transition in Eq.~(\ref{wf}). Instead of Eq.~(\ref{fk}), this
amplitude is now represented by the element $f_c^{11}$ of a $2\times
2$ matrix $\hat{f}_c$ defined in Eq.~(\ref{fc}). We have further
considered the correction $\Delta C_{K\bar{K}}$ in
Eq.~(\ref{correction}) due to the deviation of the spherical waves
from the true scattered waves in the inner region of the short-range
potential, which is of comparable size to the effect of the second
channel.

\begin{figure}[h]
\centering
\begin{minipage}[h!]{0.52\textwidth}
\centering
\includegraphics[width=\textwidth,height=0.8\textwidth]{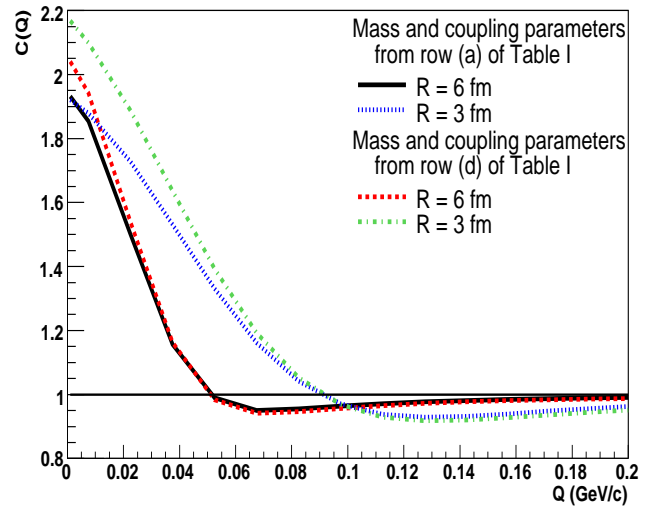}
\end{minipage}\hfill
\caption{Theoretical correlation functions for input
 Gaussian sources of $R = 6$ fm and  $R = 3$ fm with $\lambda = 1, N = 1$
The resonance masses and coupling constants are from
Table~\ref{table:TBHBT2}.
\label{ThCorrFctn}}
\end{figure}

Figure \ref{ThCorrFctn} shows the theoretical correlation functions for two sets of resonance
parameters from Table~\ref{table:TBHBT2} with $R = 6$ and $R = 3$ fm as input radii with the normalization factor $N$ and
$\lambda$ both set to unity.

The results indicate that the effect of the strong
FSI in the $K^{0} \bar{K^{0}}$ system is to give rise to
a repulsive-like component causing the correlation function to go below unity.

\section{Experimental results}
The experimental correlation functions are fit using the Lednick\'y \& Lyuboshitz \cite{LEDNICKY} model to take into account the effect of the strong FSI. The free parameters are the radius $R$ characterizing the separation $\vec{r}^*$ of the particle emission points in the pair rest frame, the normalization $N$, and $\lambda$. This fitting was done assuming the Gaussian $\vec{r}^*$--distribution of Eq.~(\ref{rDIST}).
\begin{figure}[h]
\centering
\centering
\begin{minipage}[h!]{0.52\textwidth}
\centering
\includegraphics[width=\textwidth,height=0.8\textwidth]{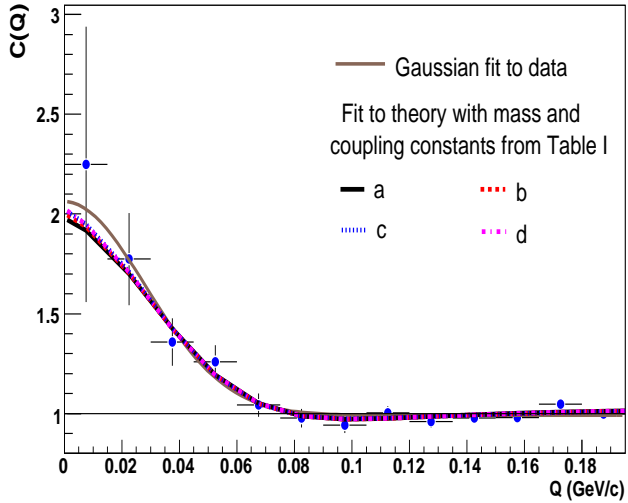}
\end{minipage}\hfill
\caption{Fits to experimental correlation function including the strong
 interaction with resonance masses and coupling constants from Table~\ref{table:TBHBT2}.
 The corresponding $\chi^2/DOF$ are (a) 1.053, (b) 1.048, (c)1.045 and (d) 1.046. 
A simple Gaussian fit, with  $\chi^2/DOF$ = 0.816, is also shown for comparison.
The errors are only statistical. 
\label{K0LednickyFits}}
\end{figure}

\begin{figure}[ht]
\centering
\begin{minipage}[h!]{0.52\textwidth}
\centering
\includegraphics[width=\textwidth,height=0.9\textwidth]{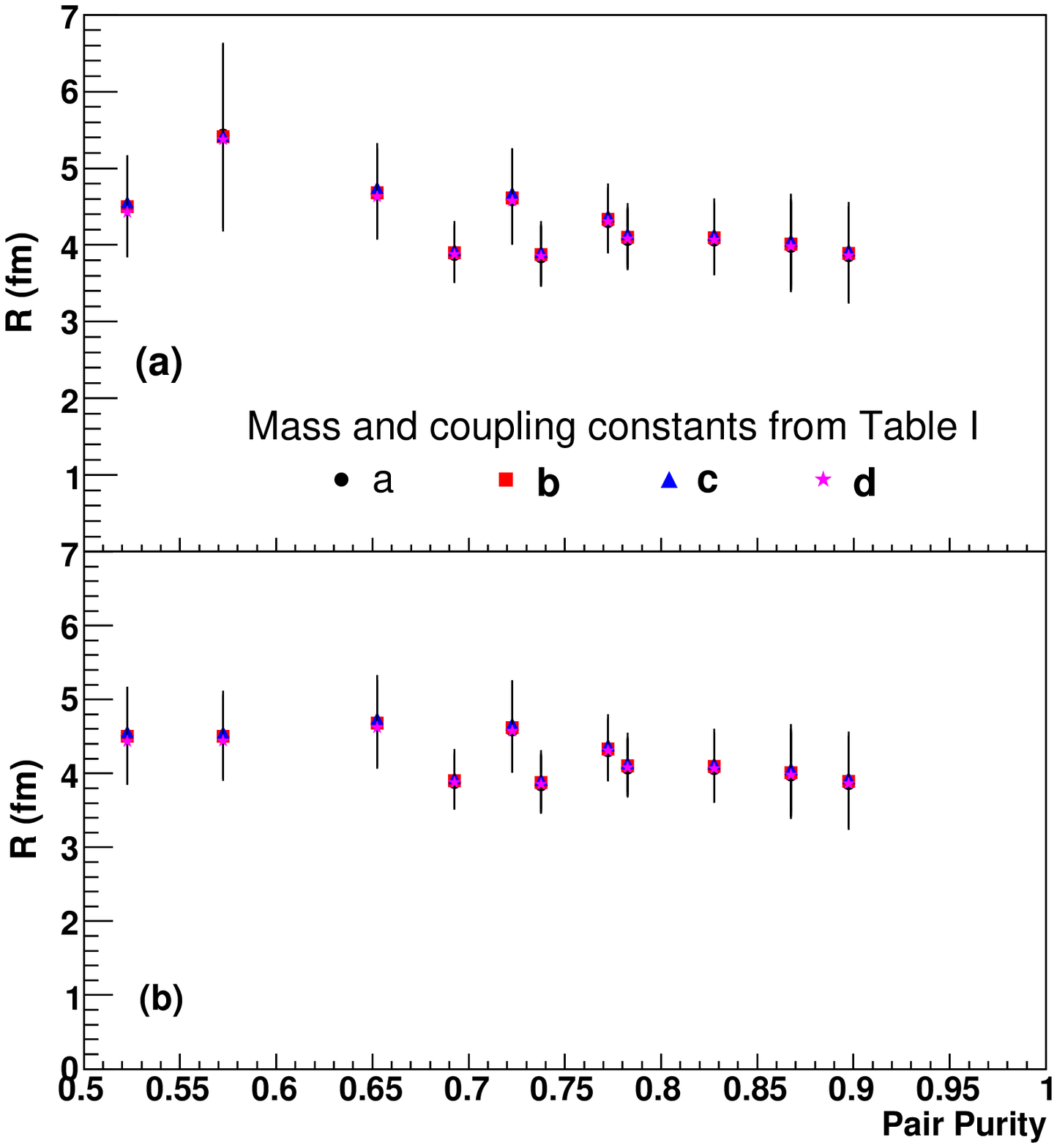}
\end{minipage}\hfill
\caption{The extracted $R$ as a function of the pair purity (a) before correction for purity and
(b) after correction for purity. The errors are only statistical.\label{RvsPairPurity}}
\end{figure}

\begin{figure}[ht]
\centering
\begin{minipage}[h!]{0.52\textwidth}
\centering
\includegraphics[width=\textwidth,height=0.9\textwidth]{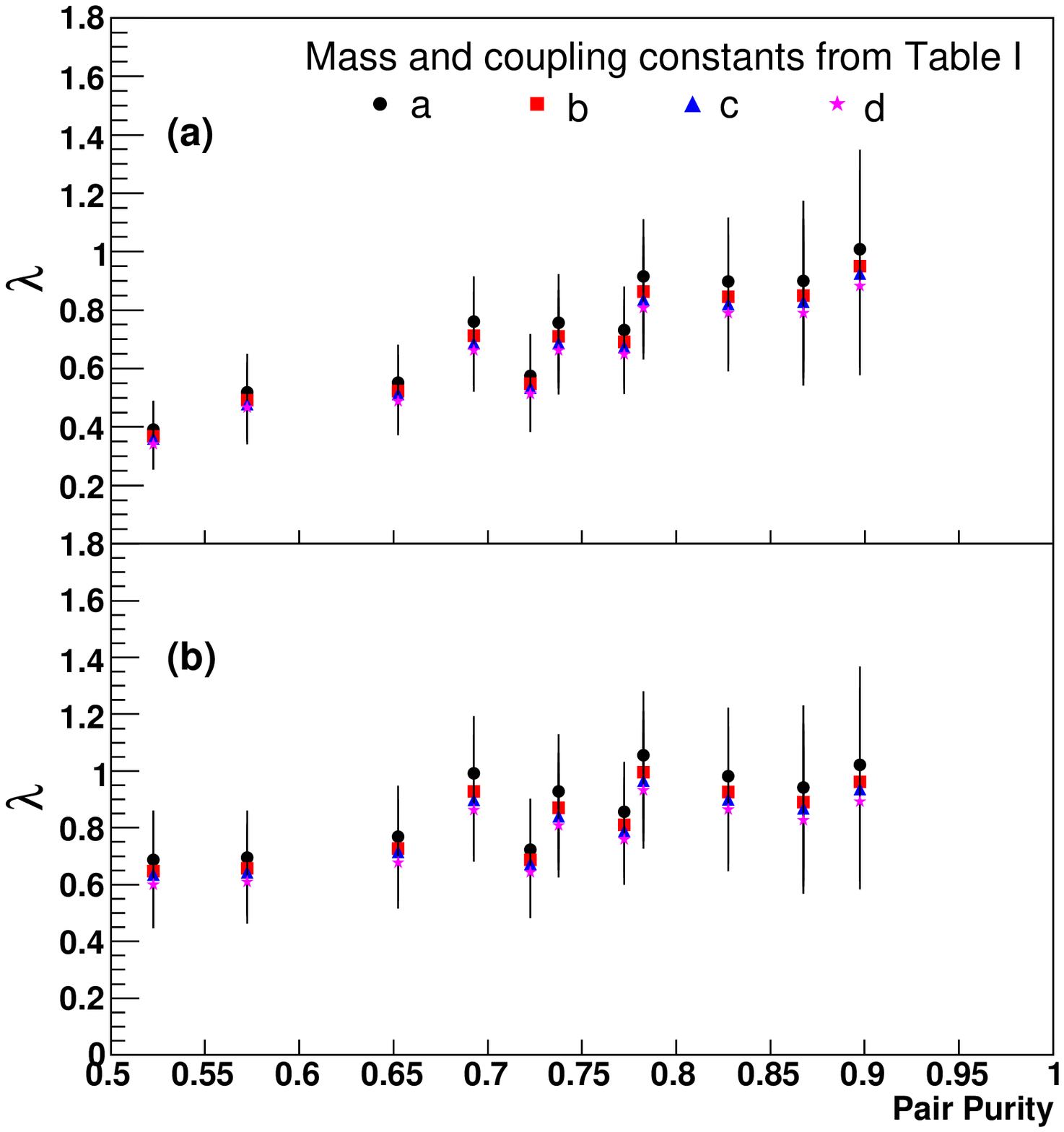}
\end{minipage}\hfill
\caption{The extracted $\lambda$ as a function of the pair purity
(a) before correction for purity and (b) after correction for
purity. The errors are only statistical.\label{LamvsPairPurity}}
\end{figure}

\begin{figure}[ht]
\centering
\begin{minipage}[h!]{0.48\textwidth}
\centering
\includegraphics[width=\textwidth,height=0.8\textwidth]{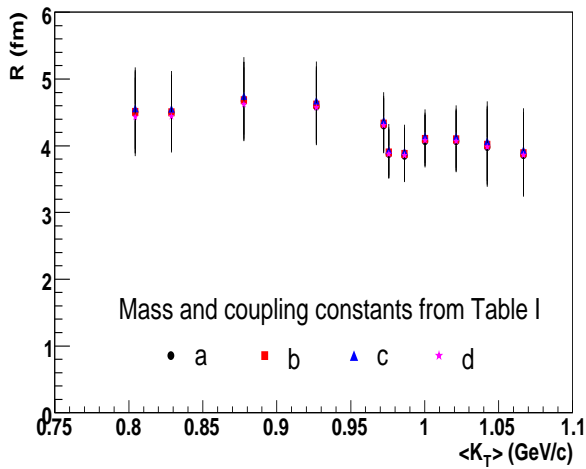}
\end{minipage}\hfill
\caption{The extracted $R$ as a function of the mean $K_{T}$ of the pairs
that go into the correlation function. The errors are only statistical.\label{RvsMeanKt}}
\end{figure}

The fit results are summarized in Table~\ref{table:TBHBT3} for
various sets of resonance parameters. The normalization $N = 1.03$
in all cases. The difference between the single channel and coupled
channel fits is very small, but it is the coupled channel fit
results which are more accurate. Figure \ref{K0LednickyFits} shows
an example of the model fits to the experimental correlation
function. A Gaussian fit to the correlation function gives 
$ R=5.02 \pm 0.61 $ fm and $\lambda = 1.08 \pm 0.29$.
One can see that a Gaussian fit cannot account for the $C(Q) < 1$
part of the data which are fit better if the strong FSI is included.
Figures \ref{RvsPairPurity} and \ref{LamvsPairPurity} show the
dependence of the extracted $R_{inv}$ and $\lambda$ parameters as a
function
  of the $PairPurity$ before, (a), and after, (b), correcting for this impurity. 
The data points are not independent of each other as a low purity data may contain
 some or all of the high purity data. The fit results are not
sensitive to the resonance parameters used. Hence, the systematic
errors are driven by the data and not theory. Figure
\ref{RvsPairPurity} shows only a slight dependence of the radius
parameter on the pair purity. On the other hand, $\lambda$ in panel (a) of Figure
\ref{LamvsPairPurity} has a strong dependence on pair purity. Even
though the purity correction seems to improve the results, there is
still a slight dependence remaining as shown in panel (b) of Figure
\ref{LamvsPairPurity}. The value of
 $\lambda$ for the data with the highest purity, and therefore the cleanest signal, is consistent with unity.
This is expected for a chaotic system with little contributions from decaying resonances. Plotting the radius
as a function of the mean $K_T$, as shown in Figure \ref{RvsMeanKt}, shows a slight dependence of $R$ with
increasing $K_T$. However this could be a remaining artifact of the mean $K_T$ dependence on pair purity, as mentioned
earlier and shown in Figure 3. One has to look at several $K_T$ bins for a specified pair purity to study
 a $K_T$ dependence of the radius coming from real physics effects. This was not possible in this analysis
 due to the limited statistics. In order to strike a balance between statistics and purity, we averaged over the
data from the coupled channel analysis corresponding the third set
of points from the right in Figure \ref{RvsPairPurity}(b), with a
pair purity of $\approx 82\%$, to obtain the values $R = 4.09 \pm
0.46 (stat.) \pm 0.31 (sys)$ fm and $\lambda = 0.92 \pm 0.23 (stat)
\pm 0.13 (sys)$ at the mean transverse  mass $\langle m_T \rangle$ = 1.07 GeV.
\begin{table}[h]
\begin{center}
\begin{tabular}{|c|c|c|}
\hline
$ ~R_{inv}$ (fm)  &  1-ch. fit  &  2-ch. fit  \\ \hline
$~a~ $ & $~  3.90 \pm 0.45 \pm 0.37 ~$ & $~ 4.07 \pm 0.46  \pm 0.31$ \\ \hline
$~b~ $ & $~  3.89 \pm 0.44 \pm 0.35 ~$ & $~ 4.09 \pm 0.46  \pm 0.31$ \\ \hline
$~c~ $ & $~  3.96 \pm 0.45 \pm 0.34 ~$ & $~ 4.14 \pm 0.47  \pm 0.31$ \\ \hline
$~d~ $ & $~  3.91 \pm 0.44 \pm 0.34 ~$ & $~ 4.07 \pm 0.45  \pm 0.29$ \\ \hline
\end{tabular}\\

\begin{tabular}{|c|c|c|c|}
\hline
$~~~~~~\lambda~~~~~~$  & 1-ch. fit & 2-ch. fit \\ \hline
$~~a~~ $ & $~  0.89 \pm 0.21 \pm 0.10~$ & $~ 0.98 \pm 0.24 \pm 0.14$ \\ \hline
$~~b~~ $ & $~  0.83 \pm 0.20 \pm 0.10~$ & $~ 0.93 \pm 0.23 \pm 0.13$ \\ \hline
$~~c~~ $ & $~  0.81 \pm 0.20 \pm 0.09~$ & $~ 0.90 \pm 0.23 \pm 0.12$ \\ \hline
$~~d~~ $ & $~  0.78 \pm 0.19 \pm 0.09~$ & $~ 0.86 \pm 0.22 \pm 0.12$ \\ \hline

\end{tabular}\\
\end{center}
\caption{The values of the radius $R$ in fm and the suppression
parameter $\lambda$ obtained by fitting the experimental correlation
function with the model \cite{LEDNICKY} that takes into account the
FSI effect in the resonance ($f_0 + a_0$) approximation. The
normalization $N = 1.03$ in all cases. The values
correspond to the third set of points from the right in Figure \ref{RvsPairPurity},
 so chosen as to strike a balance between statistics and purity.
The results in the first and the second column respectively
correspond to the single- and two-channel fits. The errors are, from
left to right, statistical and systematic errors introduced by the
uncertainty on the purity correction. The systematic errors from the
model fits are very small in comparison and are not shown.
\label{table:TBHBT3}}
\end{table}


\begin{figure}[h]
\centering
\centering
\begin{minipage}[h!]{0.48\textwidth}
\centering
\includegraphics[width=\textwidth,height=0.8\textwidth]{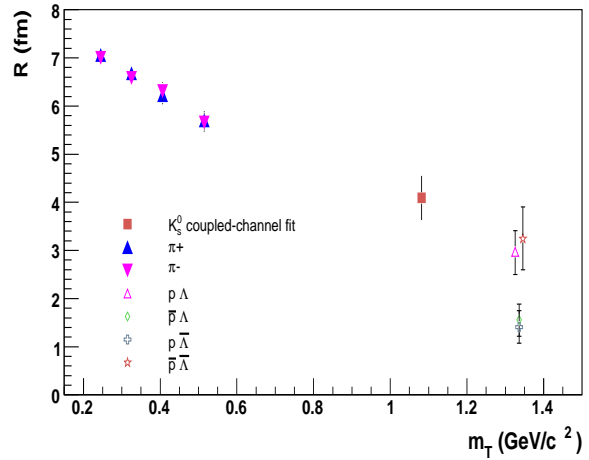}
\end{minipage}\hfill
\label{K0MtSys}
\caption{$R$ as a function of $m_T$.
Statistical and systematic errors are shown. The FSI uncertainty measured by the
spread of the fit results in rows (a)-(d) of Table~\ref{table:TBHBT3} is substantially smaller
than the statistical error.
\label{RinVsRout}\label{MtPLOT}}
\end{figure}

Figure \ref{MtPLOT} shows the $m_T$ dependence of $R$ extracted from
$\pi\pi$ \cite{MERCEDES}, $K^0_sK^0_s$, and proton-$\Lambda$ correlations \cite{GAEL}. Considering
the large mean transverse momentum of the pair, the value of
 $R$ for $K^0_s$ before taking into account the FSI in  the $K^0\bar{K^0}$
system is larger than expected from the systematics followed by the
rest of the data.  However, after taking into account the
FSI effect the neutral kaons also seem to follow the $m_T$ scaling that
hydrodynamics predicts \cite{HEINZ96}.

\section{Conclusions}
We have presented the first measurement of neutral kaon correlations
in heavy-ion collisions at RHIC. One has to consider the effects of FSI to
obtain reasonable agreement between theory and data. The variations
of the resonance parameters result in very small differences, which
are well within our systematic errors. The effect of the pair purity
on the correlation function has been studied extensively and is well
understood. A Gaussian fit to the correlation function does account
 very well for the  $C(Q) < 1$ part of the data and gives a radius which is
larger compared to the model fit results.

The measured correlation
 radius is intermediate between those obtained from two-pion and
proton-lambda correlations in these collisions with the same
conditions  except for a different transverse mass, $m_T$. The radii
seem to follow
 a universal $m_T$ dependence in agreement with a universal collective
 flow predicted by hydrodynamics. The value of the parameter
 $\lambda$, based on the high purity data, is consistent with unity and thus points
 to a chaotic kaon source. This is in correspondence with an indication
of a dominantly chaotic pion source obtained from STAR measurement of
three pion correlations \cite{JADAMS}. 

Our results represent an important
 first step towards a multi-dimensional analysis of neutral kaon correlations
 using the high statistics data from RHIC. In the future this analysis will allow to extract
information about the freeze-out geometry, collective flow velocity,
 the evolution time and duration of particle emission. The latter is especially
 interesting in the context of an increased emission duration expected if there is a
first order phase transition from a quark gluon plasma to a hadronic
system. Recent pion interferometry measurements at RHIC however
point to a smaller evolution time and emission duration than expected
from the usual hydrodynamic and transport models. This result may
indicate an explosive character of the collision and is often
considered as the interferometry puzzle. The fact that the Coulomb
interaction is absent in the dominant elastic transition and that
the FSI effect can be handled with sufficient accuracy makes neutral
kaon interferometry a powerful tool which allows for an important
cross-check of charged pion correlation measurements. Pion
measurements are much more strongly affected by contributions from
resonance decays and final state interactions.

\section{Acknowledgements}

We thank the RHIC Operations Group and RCF at BNL, and the
NERSC Center at LBNL for their support. This work was supported
in part by the Offices of NP and HEP within the U.S. DOE Office 
of Science; the U.S. NSF; the BMBF of Germany; CNRS/IN2P3, RA, RPL, and
EMN of France; EPSRC of the United Kingdom; FAPESP of Brazil;
the Russian Ministry of Science and Technology; the Ministry of
Education and the NNSFC of China; IRP and GA of the Czech Republic,
FOM of the Netherlands, DAE, DST, and CSIR of the Government
of India; Swiss NSF; the Polish State Committee for Scientific 
Research; SRDA of Slovakia, and the Korea Sci. \& Eng. Foundation.

\section{Appendix}

The interaction of final state particles can proceed not
only through the elastic transition $ab \rightarrow ab$ but also
through inelastic reactions of the type $cd \to ab$, where c
and d are also final state particles of the production process. The FSI effect
 on particle correlations is known to be significant
 only for particles with a slow relative motion. Such particles continue to 
interact with each other after leaving the domain of particle production and their
slow relative motion guarantees the possibility of the separation (factorization)
 of the amplitude of a slow FSI from the amplitude of a fast production process.
For the relative motion of the particles involved in the FSI to be slow, 
the sums of the particle masses in the entrance and exit channels should be close to each
other \cite{LLL}. Thus, in our case, one should account for the effect of inelastic 
transition  $K^+K^- \rightarrow K^0\bar{K^0}$ in addition to the elastic
 transition $K^0\bar{K^0}  \rightarrow K^0\bar{K^0}$. 
Instead of a single channel Scr\"odinger equation one should thus solve a two-channel one.
In solving the standard scattering problem, one should take
into account that the FSI problem corresponds to the inverse direction
of time. As a result, one has to make the substitution  $\vec{k}^* \to -\vec{k}^*$ and consider 
$K^0\bar{K^0}(\equiv 1)$ as the entrance channel and $K^+K^- (\equiv 2)$ as the exit channel.
 Since the particles in both channels are members of the same isospin
multiplets, one can assume that they are produced with about the
same probability. Therefore the correlation function will be simply
a sum of the average squares of the wave functions
$\Psi^{11}_{-\vec{k}^*}(\vec{r}^*)$ and
$\Psi^{21}_{-\vec{k}^*}(\vec{r}^*)$ describing the elastic and
inelastic transitions respectively.

Assuming the s-wave dominance and $r^*$ outside the range
of the strong interaction potential, one has \cite{LLL}:
\begin{equation}
\label{wf21}
\Psi^{21}_{-\vec{k}^*}(\vec{r}^*)=f^{21}_c(k^*)
\sqrt{\frac{\mu_2}{\mu_1}} \frac{\tilde{G}(\rho_2,\eta_2)}{r^*},
\end{equation}
where $\mu_1=m_{K^0}/2$ and $\mu_2=m_{K^+}/2$ are the respective
reduced masses in the two channels. $\rho_2=k_2^*r^*$,
$\eta_2=(k_2^*a_2)^{-1}$ and
$k_2^*=[2\mu_2(k^{*2}/(2\mu_1)+2m_{K^0}-2m_{K^+}]^{1/2}$ is the
$K^+$ momentum in the two-kaon rest frame. $a_2=-(\mu_2 e^2)^{-1}= -
109.6$ fm is the (negative) $K^+K^-$ Bohr radius, $f^{21}_c(k^*)$ is
the s-wave transition amplitude re-normalized by Coulomb interaction
in the $K^+K^-$ channel, $\tilde{G}(\rho,\eta)=\sqrt{A_c(\eta)}.
[G_0(\rho,\eta)+iF_0(\rho,\eta)]$ is the combination of the singular
and regular s-wave Coulomb functions $G_0$ and $F_0$. Finally
$A_c(\eta)=2\pi\eta/[\exp(2\pi\eta)-1]$ is the Coulomb penetration
(Gamow) factor.

The wave function of
the elastic transition $1\to 1$ is still given by
Eq.~(\ref{wf}) in
which $k^*\equiv k_1^*$ and the amplitude $f=f_c^{11}$
is now the
element of a $2\times 2$ matrix
\begin{equation}
\label{fc}
\hat{f}_c=\left(\hat{K}^{-1}-i\hat{k}_c\right)^{-1}.
\end{equation}
Here $\hat{K}$ is a symmetric matrix and $\hat{k}_c$ is a diagonal
matrix in the channel representation: $k_c^{11}=k^*$,
$k_c^{22}=A_c(\eta_2)k_2^*-2ih(\eta_2)/a_2$, where the function
$h(\eta)$ is expressed through the digamma function
$\psi(z)=\Gamma'(z)/\Gamma(z)$ as
$h(\eta)=[\psi(i\eta)-\psi(-i\eta)-\ln\eta^2]/2$. Assuming that the
isospin violation arises solely from the mass difference and Coulomb
effects on the element $k_c^{22}$, making it different from the
momentum $k^*$ in the neutral kaon channel, one can express the
$\hat{K}^{-1}$ matrix, in the channel representation through the
inverse diagonal elements $K_I^{-1}$ of the $\hat{K}$-matrix in the
representation of total isospin $I$ (the products of the
corresponding Clebsch-Gordan coefficients being 1/2 or -1/2):
\begin{eqnarray}
\label{isospin}
(\hat{K}^{-1})^{11}=(\hat{K}^{-1})^{22}=
\frac12\left[{K}^{-1}_{0}+{K}^{-1}_{1}\right],
\nonumber\\
(\hat{K}^{-1})^{21}=(\hat{K}^{-1})^{12}=
\frac12\left[{K}^{-1}_{0}-{K}^{-1}_{1}\right].
\end{eqnarray}
The latter are assumed to be dominated by the resonances
$r=f_0(980)$ and $a_0(980)$ for $I=0$ and 1, respectively, so:
\begin{equation}
\label{Kreson}
{K}^{-1}_I=(m_r^2-s-ik_r'\gamma_r')/\gamma_r .
\end{equation}

One should also take into account the correction $\Delta
C_{K\bar{K}}$ due to the deviation of the spherical waves from the
true scattered waves in the inner region of the short-range
potential, which is of comparable size to the effect of the second
channel. This correction is also given in Ref. \cite{LLL} and is
represented in a compact form in Eq.~(125) of Ref.~\cite{L05}. In
our case,
\begin{eqnarray}
\label{correction}
\Delta C_{K\bar{K}}&=&-\frac{1}{4\sqrt{\pi}R^3}
\left[|f_c^{11}|^2d_0^{11}+|f_c^{11}|^2d_0^{11}\right.
\nonumber\\
& &+\left. 2\Re(f_c^{11}f_c^{21*})d_0^{21}\right],
\end{eqnarray}
where $d_0^{ij}=2\Re d(\hat{K}^{-1})^{ij}/dk^{*2}$;
at $k^*=0$, $\hat{d}_0$ coincides with the real part of
the matrix of effective radii.

One may see from Eqs.~(\ref{cft}) and (\ref{wf21})
that the usual resonance Breit-Wigner behavior settles only
at small $r^*$ when
squares of the spherical waves $|f_c^{ij}/r^*|^2$ dominate.
At sufficiently large $k^*$, one can neglect the Coulomb effects
and put $f_c^{11}\doteq (f_0+f_1)/2$, $f_c^{21}\doteq (f_0-f_1)/2$,
so that $|f_c^{11}|^2+|f_c^{21}|^2\doteq|f_0|^2+|f_1|^2$. The
sum of the square terms then reduces to the incoherent
Breit-Wigner contributions of $f_0$ and $a_0$ resonances.
There can also be additional (not related to FSI)
resonance contribution of the usual Briet-Wigner form due to
direct $f_0(980)$ and $a_0(980)$ production. This
contribution is assumed to be negligible as compared to the
FSI effect.

\end{document}